\documentstyle[twocolumn,prl,aps,ams,epsf]{revtex}
\newcommand{\abst}{ }
\begin{document}
\draft
\title{Stochastic Resonance in Chaotic Spin--Wave Dynamics}
\author{Ekkehard Reibold\thanks{e--mail: 
reibold@exp1.fkp.physik.th-darmstadt.de}, 
Wolfram Just\thanks{e--mail: 
wolfram@arnold.fkp.physik.th-darmstadt.de}, Jens Becker,
and Hartmut Benner\thanks{e--mail: 
benner@hrzpub.th-darmstadt.de}}
\address{ Institut f\"ur Festk\"orperphysik, Technische Hochschule Darmstadt,
Hochschulstra\ss e 6, D--64289  Darmstadt, Germany}
\date{January 31, 1997}
\maketitle 
\begin{abstract}
{ We report the first experimental observation 
of {\it noise--free} stochastic resonance 
by utilizing the intrinsic chaotic 
dynamics of the system.
To this end we have investigated the effect of an
external periodic modulation on intermittent signals 
observed by high power ferromagnetic resonance in
yttrium iron garnet spheres.}
Both the signal--to--noise ratio and the residence 
time distributions show the characteristic features of stochastic resonance.
The phenomena can be explained by means
of a one--dimensional intermittent map. We present analytical results 
as well as computer simulations.
\end{abstract}
\section{Introduction}
Stochastic resonance, invented 15 years ago as a model for geophysical
dynamics \cite{Ben81}, has meanwhile found its way into such diverse fields
like physics, meteorology, chemistry, and biology \cite{Mos93,Mos94,Man95}. 
The rising interest in this field stems
from the counterintuitive effect that a periodic signal component 
can be amplified by a stochastic force. 

The basic mechanism is usually
explained { in terms of the famous Kramers problem \cite{Kra40}, 
i.e.~the overdamped motion of a particle in a
symmetric double--well potential subjected to noise, which is
supplemented by a time periodic forcing.
The noise causes ''incoherent tunnelling'' 
between the two wells with an exponentially decreasing
distribution of the respective residence times.
The periodic forcing leads to enhanced 
transitions on certain time scales and, therefore, to a periodic 
signal component. It is the prominent feature
of stochastic resonance that the signal-to-noise ratio 
does not decrease with increasing noise amplitude, but}
attains a maximum at a certain noise strength.
A second characteristic property shows up in
the distribution of residence times, 
{ where} the periodic forcing leads to maxima at odd 
multiples of half the driving period (cf.\cite{Zho90}). 
Of course, these signatures of stochastic resonance { are not confined
to this special model, but} occur in more
general bi-- and monostable systems and for different types of 
noise \cite{Mos93,Man95}. 

{ It was suggested by Anishchenko et al.~\cite{Ani93} based on
the numerical analysis of a simple map, that quite similar
phenomena can be caused by the internal chaotic dynamics instead of
an external noise. In that case the intermittent hopping between
different chaotic repellers in conjunction with an additional
periodic forcing can be used to amplify the periodic signal component
in close analogy to conventional stochastic resonance.}
This mathematical model system motivated us to look for
{\it noise--free} stochastic resonance in real experimental systems.

In this letter we report on its realization in chaotic spin--wave dynamics.
By taking advantage of the known bifurcation scenarios beyond
the instability threshold \cite{Roed,bnrconf}, we have
investigated the chaotic intermittent dynamics subjected to an
additional periodic forcing.
\section{Experiments and results}
High power ferromagnetic resonance 
experiments were performed on spheres of yttrium iron garnet (YIG), 
which is well established as a ''prototype nonlinear ferromagnet''
and has extensively been studied for decades \cite{Dam,Ben92,Wig94}.
The sample was placed in a bimodal 
transmission--type microwave 
cavity and excited by a microwave field of $9.25\,\mbox{GHz}$,
applied perpendicularly to the static magnetic field {\boldmath$H$}.
The parametric excitation of spin waves was observed in
{\it subsidiary absorption} \cite{Suh57}.
The transmitted and rectified signal was 
recorded with a digital transient recorder and analysed on a PC. 
Accessible system parameters are the static 
magnetic field and the microwave power, which is proportional to the 
squared amplitude of the pumping field. 
Increasing the pumping power above
the first--order 
Suhl threshold \cite{Suh57} we observed auto--oscillations and a variety of
routes into chaos: period doubling, quasi--periodicity, and 
different types of intermittency 
\cite{Ben92,Wig94}. Our measurements were performed in a 
parameter region of type--III intermittency, { where 
the system jumps between period 2 behaviour (regular phase) and
chaotic behaviour (chaotic phase).}
The intermittency scenario can be run through in 
two ways, either by varying the microwave
input power $P$ or the static magnetic field 
$H$ \cite{bnrconf}. 

Here, the regular and { chaotic} phases take the role of the two
states in conventional models of stochastic resonance, and the
intrinsic chaotic dynamics acts as a ''stochastic'' forcing.
Accordingly, the mean lengths of the two phases
can be changed on variation of the two external control parameters
$P$ and $H$. In this
sense $P$ and $H$ correspond to the noise strength in conventional
stochastic resonance experiments. 
To obtain the resonance effect
{ one has} to apply an additional periodic forcing, which in our 
case consists 
of an amplitude modulation of the microwave power. The two
additional parameters, modulation frequency and
amplitude, have to be adjusted properly, i.e.~the 
modulation amplitude has to be 
chosen small enough to ensure that the system stays inside the
intermittency regime, but large enough 
to observe a periodic component in the transmitted signal.
\begin{figure}
\epsffile{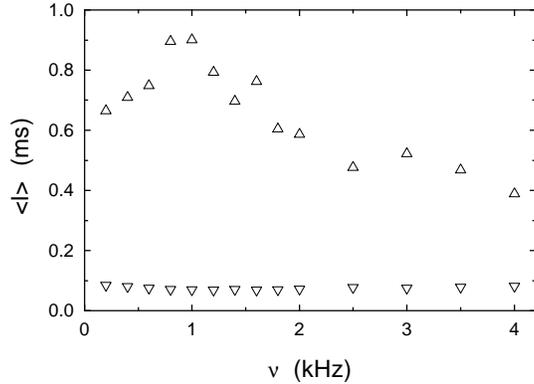} \abst
\caption{Mean lengths of the chaotic $( \triangle )$ and the regular
$( \nabla )$ phases vs.~modulation frequency.}
\label{fig:in1}
\end{figure}
\begin{figure}
\epsffile{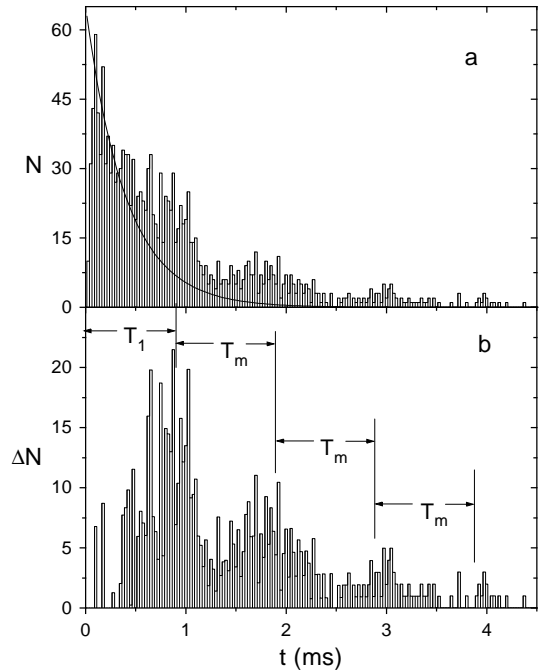} \abst
\caption{(a) Probability distribution of the chaotic lengths 
for $\nu = \nu_{max}$. 
The full line indicates
the exponential background. (b) Difference between 
the total probability 
and the exponential background taken from (a).}
\label{fig:in2}
\end{figure}
The influence of modulation frequency shows up in the { duration} 
of regular and chaotic phases (cf.~fig.\ref{fig:in1}). 
We found that the mean length of the regular state
is not affected by the modulation. 
The mean chaotic length shows a distinct maximum 
at $\nu_{max}=1 \mbox{kHz}$, which means that
the reinjection from the chaotic to the regular 
state becomes less probable. 
At $\nu_{max}$ the mean return time, i.e.~the 
sum of the two lengths, is of the order of the modulation period,
which is a prerequisite for the occurrence of stochastic resonance.

Signatures of stochastic resonance could be observed in the
{ distribution of the durations of the chaotic phase, 
i.e. the residence time distribution.}
Without modulation one would obtain an exponential decay, caused by
the uniform reinjection from the chaotic to the regular state.
With modulation there appears a structure 
on top of this background, which represents
a typical feature of stochastic resonance \cite{Zho90}. 
The distribution at $\nu_{max}$
is plotted in fig.~\ref{fig:in2}(a). 
The exponential background resulting from the intermittent dynamics
is larger than in conventional stochastic resonance. 
Figure \ref{fig:in2}(b) shows the distribution after
having subtracted this background.
The distance between the peaks is equal to the 
modulation period $T_m$, as expected from theory.
Since our system is strongly asymmetric the first peak is 
not located at exactly half the period but shifted towards a higher value. 

\begin{figure}
\epsffile{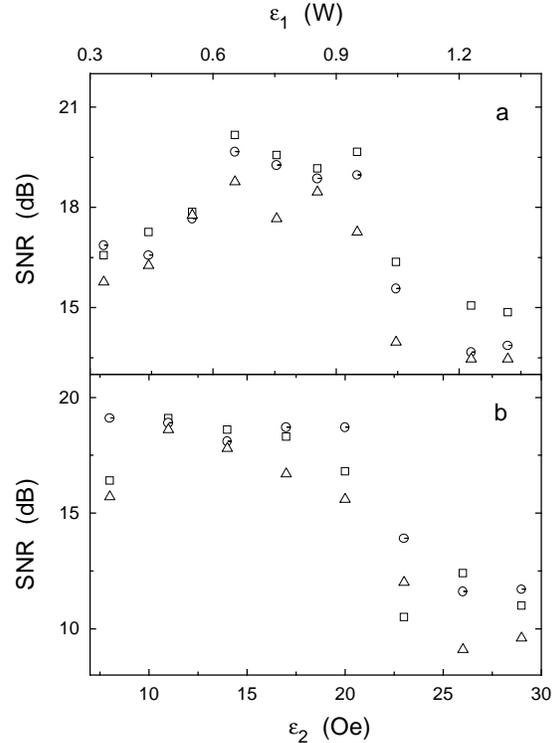} \abst
\caption{Signal--to--noise ratio vs.~the bifurcation parameters 
$\varepsilon_1 =P_c-P$ (a) and $\varepsilon_2 = H_c - H$ (b) for various 
modulation frequencies ( $\Box\, 1.0 \mbox{kHz},$ 
$\circ\, 5.0 \mbox{kHz},$
$\triangle\, 10.0 \mbox{kHz}$ ).}
\label{fig:in3}
\end{figure}
Stochastic resonance was originally defined by an increase of the 
signal--to--noise ratio with increasing noise strength. 
In our experiments we 
have two possibilities to change the 
internal ''deterministic noise level''. 
At the bifurcation point ($H=H_c$, $P=P_c$)
the regular state is marginally stable. With increasing 
bifurcation parameters 
$\varepsilon_1 =P_c-P$ and  $\varepsilon_2 = H_c - H$ \cite{bnrconf}
the switching rate between regular and 
chaotic states increases and hence the internal noise level, too. 
If we take, as usual, the amplitude ratio between peak 
and background of the Fourier spectrum as a definition 
of the signal--to--noise ratio,
we observe a maximum on variation of the static magnetic field or 
the microwave power.
Figure \ref{fig:in3} shows the results for several modulation
frequencies on a logarithmic scale. Distinct resonance phenomena 
can be seen in both the field and power dependences. 
\section{Theoretical model and simulations}
\begin{figure}
\epsffile{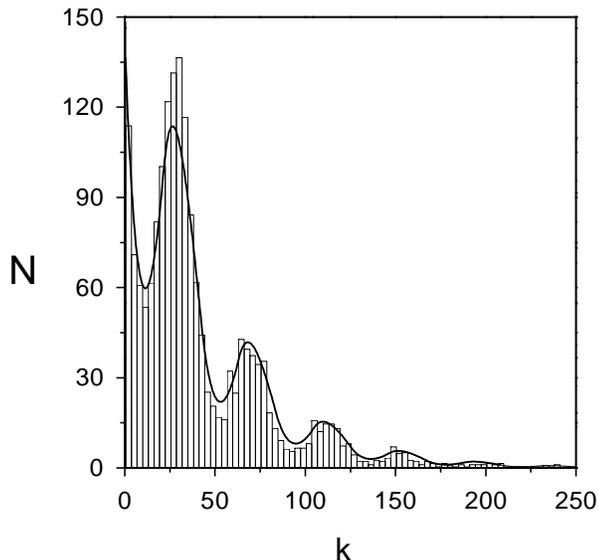} \abst
\caption{Distribution of chaotic lengths. Analytical result (solid line)
and simulation.}
\label{fig:in5}
\end{figure}

{ Far above the Suhl threshold an approach from first principles
has turned out to be extremely 
complicated \cite{Ben92}. Thus we adopt a phenomenological
description in terms of simple mathematical model systems, which
have proven to be fruitful in diverse
problems of nonlinear dynamics.}
In order to develop a theoretical explanation for the residence time
distribution we propose a time
dependent one--dimensional map which can be treated analytically.
\begin{figure}
\epsffile{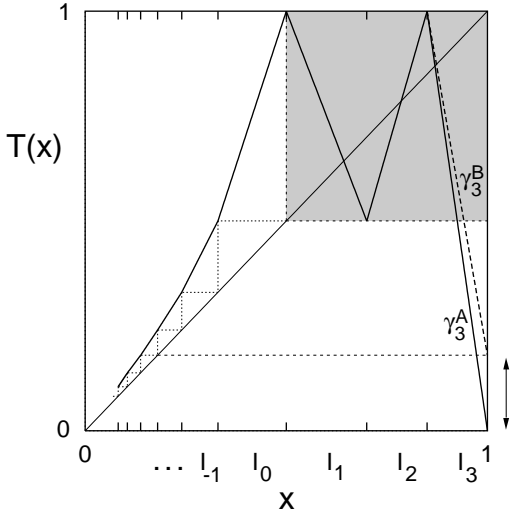} \abst
\caption{{ Piecewise linear map} $x_{n+1}=T(x_n)$. 
The broken line in $I_3$ { and the double arrow
show} the time dependent modulation.
The { shaded} box in the upper right corner
indicates the region of the chaotic repeller.}
\label{fig:in4}
\end{figure}
{ This map generates a regular phase via an inverse
pitchfork bifurcation and a chaotic phase via a chaotic repeller
(cf.~fig.\ref{fig:in4}).} 
We adopt a piecewise linear model with slopes
$\gamma_l$ on the intervals $I_l$
to keep the analysis entirely analytical \cite{Mori86}. 
A periodic modulation with period $n_p$ is included
via the time dependence of the slope in the right--hand interval $I_3$.
For simplicity we confine ourself to the case where
the slope $\gamma_3$ changes
between two different values $\gamma_3^A$ and $\gamma_3^B$ every
$(n_p/2)^{th}$ time step. 
{ Let $\rho_{l}(n)$ denote the probability density for a trajectory
hitting the interval $I_l$ at time $n$. It is periodic in time 
with period $n_p$ but 
attains a constant value on each interval, since the model maps 
intervals on intervals (Markov map). Furthermore let}
$S_k(n)\subseteq I_0$ denote those
phase space points which stay exactly $k$ time steps in the chaotic
region $I_1\cup I_2\cup I_3$ if the initial phase of the {
modulation} is $n$,
{ and denote the size of this set by $\lambda(S_k(n))$.}
Then $\rho_0(n) \lambda(S_k(n))$ gives the probability that
at phase $n$ of the { modulation}
a chaotic burst of length $k$ occurs.
Hence by averaging over the initial phase we obtain the desired
distribution of residence times
\begin{equation}\label{aa}
N_k= \frac{1}{
n_p \bar{\rho}_0 \lambda(I_0)}
\sum_{n=0}^{n_p-1} \rho_0(n) \lambda(S_k(n)) \quad ,
\end{equation}
where $\bar{\rho}_0$ denotes the average { of $\rho_0(n)$} over
one period. The structure of the distribution is determined by the
time dependence of the density as well as by the time
dependence of $S_k(n)$. 
{ The size of $S_k(n)$ can be expressed by the (static) escape 
rates of the chaotic repeller \cite{Rand87}. Using the abbreviation
\begin{equation}\label{ab}
p_k(n)= \left\{\begin{array}{lcl}
p_A &:=& 1/|\gamma_1|+1/|\gamma_2|+1/|\gamma_3^{A}|\\
 &\mbox{if}& (k+n)|\mbox{mod } n_p < n_p/2 \\
p_B &:=& 1/|\gamma_1|+1/|\gamma_2|+1/|\gamma_3^{B}|\\
 &\mbox{if}& (k+n)|\mbox{mod } n_p \geq n_p/2
\end{array}
\right. \quad ,
\end{equation}
a geometrical consideration yields}
\begin{equation}\label{ac}
\lambda(S_k(n)) = \lambda(I_0) \prod_{i=1}^{k-1}
p_i(n) \cdot (1-p_k(n)) \quad .
\end{equation}
The density $\rho_0(n)$ can in principle be
evaluated, too. But to leading order in the { modulation amplitude it
remains  time independent $\rho_0(n)\simeq \bar{\rho}_0$. 
This implies that the properties of the regular phase are not
affected by the modulation, which is in
accordance with our experimental observation.
With eq.(\ref{ac}) the distribution (\ref{aa}) reads}
\begin{equation}\label{ad}
N_k\simeq \frac{1}{n_p} \sum_{n=0}^{n_p-1}
\prod_{i=1}^{k-1} p_i(n) \cdot (1-p_k(n)) \quad .
\end{equation}
Its structure is entirely determined by the time dependence of 
the escape rate and does not depend on details of the intermittency
mechanism. Expression (\ref{ad}) is evaluated by combinatorics. In the
case of large period $n_p\gg 1$ the distribution becomes 
quasi--continuous
\begin{equation}\label{ae}
N_k \simeq \exp(- \sigma k /n_p) \left\{\begin{array}{ll}
f_{-}(x), & 0 \le x < 1/2\\
f_{+}(x), & 1/2 \le x < 1
\end{array}\right. \quad ,
\end{equation}
{ where $x=k/n_p |\mbox{mod } 1$ denotes the fractional part of the 
chaotic length with respect to the modulation period.
Expression (\ref{ae}) reflects an exponential decay of the distribution
with rate $\sigma:=n_p[1-(p_A+p_B)/2]$
and a superstructure of period $n_p$ 
(cf.~fig.\ref{fig:in5}), which is explicitly given by 
\begin{eqnarray}\label{af}
& & f_{\pm}(1/2\pm z) 
:= 2 z \sigma \cosh[\delta(1/2-z)]\nonumber\\
&\pm& 2 z \delta \sinh[\delta (1/2-z)]
+ 2\sigma/\delta \sinh[\delta(1/2-z)] \quad .
\end{eqnarray}
Here $\delta:=(p_B-p_A)n_p/2$ represents
the amplitude of the modulation.}

{ In addition to}
our analytical results we have performed computer simulations
using a one--dimensional map like fig.~\ref{fig:in4} but with
the left--hand part being replaced by a cubic function.
The time dependence was introduced by modulation of
the right--hand 
maximum. From the time series we evaluated the distribution of the residence
times which are shown in fig.~\ref{fig:in5}. Like for the analytical
results we found an exponential decay 
modulated periodically with the period of the forcing. 
In this case, like for the ferromagnetic resonance 
experiments, a large exponential background occurs
which is caused by the internal { chaotic dynamics, i.e.~the}
''deterministic noise level''.
The coincidence is even more convincing when keeping in mind that
the analytical prediction and the simulation have been obtained from
different maps and different modulation mechanisms.
\section{Conclusions}
We have reported on the experimental realization of stochastic 
resonance in ferromagnetic resonance experiments beyond the
{ Suhl threshold. In contrast to conventional stochastic
resonance we used
the interplay between the
intrinsic chaotic dynamics in an intermittent parameter 
region and an external periodic modulation
for the realization of {\it noise--free} stochastic resonance.
Nevertheless the signal--to--noise ratio and 
the residence time distributions show exactly the same
characteristic behaviour as in conventional stochastic resonance.}
We were able to explain these features
in terms of a one--dimensional intermittent
map. An analytical expression for the residence time distribution 
was derived and compared to results from computer simulations.
{ Both the quantitative agreement between theoretical calculations 
and simulations,
and the qualitative coincidence with the experimental result 
demonstrates the universality of these features.
We expect that {\it noise--free} stochastic resonance
will be realized in many other physical or technical applications 
in the near future.}
\section{Acknowledgements}
We thank Franz--Josef 
Elmer and Janusz Ho{\l}yst for helpful and interesting 
discussions. 
This project of SFB 185 ''Nichtlineare Dynamik''  
was partly financed 
by special funds of the Deutsche Forschungsgemeinschaft.
\end{document}